# *All-Passive Nonreciprocal Metasurface*

*Ahmed M. Mahmoud, Arthur R. Davoyan, Nader Engheta[1]*

*University of Pennsylvania*
*Department of Electrical and Systems Engineering*
*Philadelphia, Pennsylvania 19104, USA*

## *Abstract*

We introduce a systematic approach to design all-passive subwavelength high performance metasurfaces that exhibit nonreciprocal properties and achieve wave-flow isolation. Moreover we build upon those findings and propose a new paradigm for a quasi-2D metasurface that mimic the nonreciprocal property of Faraday rotation without using any magnetic or electric biasing. We envision that the proposed approaches may serve as a building block for all-passive time-reversal symmetry breaking with potential applications for future nonreciprocal systems and devices.

---

[1] To whom correspnding should be addressed, email: engheta@ee.upenn.edu

*Main Text*

Recent years have witnessed an astonishing progress in design of structures and systems that control and manipulate electromagnetic waves in desired fashions (*1–5*). Of particular interest are ultrathin structures and composites, known as metasurfaces, that exhibit electromagnetic properties not readily attainable in nature (*6–8*). The interest towards these systems has sparkled mostly due to a variety of promising applications in various ranges of electromagnetic spectrum: metasurfaces are at the forefront of today's microwave technology, where compact antennas, ultrathin layers with extreme chirality and asymmetric transmission, have been proposed and fabricated (*9–11*); at optical frequencies metasurfaces are the key towards the next generation of nanophotonics with various potential applications (*6, 7, 12–14*); and even the THz and graphene-based technology is widely exploiting the metasurface principles (*15, 16*).

Metasurfaces are usually constituted of regular metal and dielectric materials specifically crafted to give a desired electromagnetic response. It is the combination of the shape and dispersion properties of the inclusions that determines the properties of the whole system. Exploiting materials with nontrivial electromagnetic properties may significantly enhance the functionality and introduce new degrees of freedom for unprecedented features. For instance, gyrotropic materials that are sensitive to the magnetic biasing may be employed to tune and control the transmission properties of the system. More specifically, metastructures with gyrotropic properties exhibit time reversal asymmetry and may be exploited for nonreciprocal propagation and transmission, e.g. Faraday rotation of light polarization and asymmetric transmission for opposite directions of illumination – the latter is a property that is crucial for the electromagnetic wave isolation (*17–22*). The ability to break the transmission symmetry is however not limited to the use of gyortropic materials; it is also possible by utilizing nonlinear material response in asymmetric systems, which have been used to design optical analogues of electronic diodes (*23–25*) in addition to using time dependent properties to break the system's reciprocity both in optical (*26*) and acoustic domains (*27*). However, the aforementioned scenarios may lead to relatively bulky and inefficient device designs. In this context developing metasurfaces that acquire efficient nonreciprocal behavior is of crucial importance. Recently metasurfaces with active electronic elements i.e. transistors, have been proposed and tested for electrically controllable more efficient nonreciprocal behavior (*28–30*). The complexities and limitations in the design and implementation of active elements and their biasing networks do, however, pose an additional complex step in such designs. Such complexities may need to be reduced in order to simplify future designs.

In this Letter we propose a concept for all-passive, not-externally-biased, subwavelength nonreciprocal metasurfaces. Our technique, which may achieve wave-flow isolation using an all passive subwavelength structures, is rather generic and in principle not limited to a specific range of operating frequency. Moreover building upon this approach, we propose a new topology where a quasi-2D metasurface can be used to mimic the non-reciprocal Faraday rotation phenomenon exhibiting very high effective magnetic activity.

Figure 1 shows the proposed concept. The metasurface unit cells consist of an all passive electromagnetic "diode" (transmitting waves in one given direction only, which we will discuss later here) and a chiral element. We design the metasurface in a checkerboard fashion considering that two neighboring unit cells are mirror images of each other. Note that chiral elements in this case lead to the polarization rotation in the opposite directions, see Fig.1. Consider illuminating the structure from one side. In this case only the unit cells with electromagnetic diodes allowing wave propagation would allow the wave to pass through. The transmitted wave would interact with the chiral structure in these cells and acquire a certain degree of polarization rotation (clockwise or counter-clockwise depending on the orientation of the chiral element in those units). On the other hand, illuminating the structure from the other side, the other set of unit "diode" cells will let the wave go through, but in these cells the chiral elements are the mirror-image of the other chiral element, thus leading to the rotation of the wave polarization in the same direction, which contrasts sharply with conventional chiral metamaterial structures and metasurfaces. Hence such a system would be mimicking the Faraday rotation of polarization. Clearly, the main challenge here is the design of an all passive subwavelength electromagnetic diode with significantly high performance characteristics. This is discussed in the next section.

It was speculated lately that by combining optical nonlinearity with a structural asymmetry it is possible to achieve a significant difference in the transmission coefficients for forward and backward propagating waves (*23*, *24*, *31*). Conventional methods for optical isolation within asymmetric nonlinear systems are based on the idea of the resonance shift in structures with distributed nonlinearities. This technique has provided up to 100 times contrast ratios between the transmission coefficients for forward and backward waves (*24*). However, to the best of our knowledge, the absolute (not relative) transmission coefficient in the desired direction of propagation has remained very low (e.g., not exceeding 0.01) in such techniques. As we demonstrate numerically below, in our technique we can overcome this limitation, while achieving wave-flow isolation with nonlinear subwavelength structures and sustaining relatively high transmission coefficient for the forward propagation direction (i.e., allowed direction in such a diode).

Here without loss of generality and for the sake of simplicity in the proof of concept we study wave-flow isolation in a rectangular waveguide at microwave frequencies. The corresponding geometry is shown in Fig. 2(a) where we have a metallic waveguide with a squared cross section 0.54 $\lambda_0$ x 0.54 $\lambda_0$, where $\lambda_0$ is the free space wavelength, that is loaded with two dielectric slabs with relative permittivities $\varepsilon_1 = 10$ and $\varepsilon_2 = 2$ , and thicknesses, $t_1 = 1.03\lambda_0$ and $t_2 = 2.03\lambda_0$. The schematic of our nonlinear subwavelength resonant structure is shown in the inset to Fig. 2(a). We implement here the principles of nonlinear metamaterial design and consider two concentric rings with four nonlinear intensity dependent capacitive elements (commonly known as varactors) (*32–34*) symmetrically placed between the rings, as shown in the inset of Fig. 2(a) considering that the nonlinear element is essentially subwavelength ring resonator (we note that our design can be easily extended to an infinite array of resonators, but here we first consider a ring resonator in a waveguide). Consider illuminating the system in the forward (+z) and backward (-z) directions. Illuminating the slab from either side causes formation of partial standing wave patterns in the structure. Owing to geometrical asymmetry of the system the field profiles inside the slab would be generally different for opposite directions of illumination, see Fig. 2 (b). Note that in linear case due to the structure's reciprocity in addition to being lossless, transmission characteristics for both directions of illumination would be identical. However, the local field intensities for forward and backward directions of illumination are essentially different and as depicted in Fig. 2(b) there exists positions along the structure where the local field ratio between the cases of forward and backward propagating waves is maximum, namely maximum local field ratio (MLFR) locations. Therefore, nonlinear, intensity dependent, resonant structures with thicknesses much smaller than the wavelength, i.e. essentially subwavelength, when placed at those locations respond differently to the two different directions of illumination. For the structure's parameters mentioned earlier we get almost about 7.5 times difference in the intensities of forward and backward directions of propagation with about 0.42 transmission coefficient as shown in Fig. 2(c).

We choose the dimensions of the resonator such that in its linear state its resonant frequency is detuned from a given operating frequency, implying that the structure is effectively transparent. For higher intensities we expect a shift of a resonance frequency towards the operating frequency with a corresponding decrease in transmission (when the resonant frequency coincides with the operating frequency almost total reflection is expected). Consequently, by inserting such a nonlinear element at the MLFR location it is possible to tune the transmission properties with respect to the direction of illumination using simply a nonlinear subwavelength resonant structure sandwiched between two thin layers, namely a substrate and a superstrate as shown in Fig. 3(a). In Fig. 3(b), we illuminate the structure with low power of 10 dBm from both sides. It is clear that for such low power, the varactors are operating in their linear regime having almost the same value of capacitance and that there is only a very slight shift between the resonance frequencies positions for forward and backward propagating waves, and consequently there is no pronounced difference in the transmission at the operating frequency. On the other hand, in Fig. 3(c) the structure is excited with a power of 30 dBm. The shift in the resonance between forward and backward propagating waves is significant, which leads to a major shift in the resonance frequencies and thus a substantial difference in the transmission coefficients for forward and backward propagation at the operating frequency. We solve the nonlinear problem numerically using CST Microwave Studio® in a steady state approximation. Fig. 3(d) shows the calculated transmission coefficients for forward and backward directions of illumination at different incident power levels. We observe that for low power levels, the transmission properties for both directions of illumination are practically the same, and the structure transmits 45% of incident radiation for both cases symmetrically. With the increase of the incident power level we observe that a transmission for backward propagating wave is decreasing, whereas the transmission in a forward direction is practically not changing. For a 30dBm incident power we already observe about 0.42 transmission coefficient for the forward illumination and about 0.04 transmission coefficient for the backward case. Moreover, the structure is robust towards changing the field polarization-angle. This stability of the structure characteristics towards the change in the incident polarization is due to the symmetry of both the resonator structure, and the distribution of the nonlinear elements around the structure.

Having a building block that behaves as an electromagnetic diode exhibiting the intended high performance all passive wave-flow isolation, we can proceed to investigating the intended all-passive quasi-2D metasurface that mimics the nonreciprocal behavior of Faraday rotation. For the sake of numerical proof of concept we choose the chiral structure in our unit cells to be the subwavelength bilayered chiral structures, allowing for almost $90^o$ polarization rotation, with almost no reflection as proposed in (*35*). A schematic of the unit cell constituting the metasurface (with two sub-unit cells as discussed before) is shown in Fig. 4(a). Fig. 4(b) shows a cross section of the electric field intensity distribution within the quasi- 2D metasurface when illuminated with a plane wave that is linearly polarized along the x-direction and is propagating in the forward (+z) direction. Only the unit cells with the electromagnetic diode oriented to allow wave propagation in the (+z) direction are excited. As depicted in Fig. 4(c) the output in that case is a rotated version of the incident field. Illuminating the surface with a plane wave propagating in the opposite (i.e., backward (–z)) direction instead, only the unit cells with the electromagnetic diodes oriented such that to allow wave propagation in the (–z) direction are excited as shown in Fig 4 (d). Since the chiral elements in these cells are mirror-image of the other chiral elements, we still get a rotated version of the incident wave, but with the same sense of rotation as the one obtained illuminating the surface from the other side as shown in Fig 4(e), mimicking the nonreciprocal Faraday rotation as required while no electric or magnetic bias is used.

In conclusion, we proposed a platform for an all-passive subwavelength metasurface with broken time-reversal symmetry. We introduced a systematic design procedure for achieving wave-flow isolation without the usage of any sort of bias (neither magnetic nor electric) that is in principle applicable to any frequency regime. Moreover we proposed a quasi-2D metasurface that mimics the non-reciprocal Faraday rotation phenomenon exhibiting very high optical activity.

This work is supported in part by the US Air Force Office of Scientific Research (AFOSR) Grant Number FA9550-10-1-0408. Some preliminary parts of this work were presented orally at the 2013 IEEE Antennas and Propagation Society International Symposium, Ref [8].

# Figures

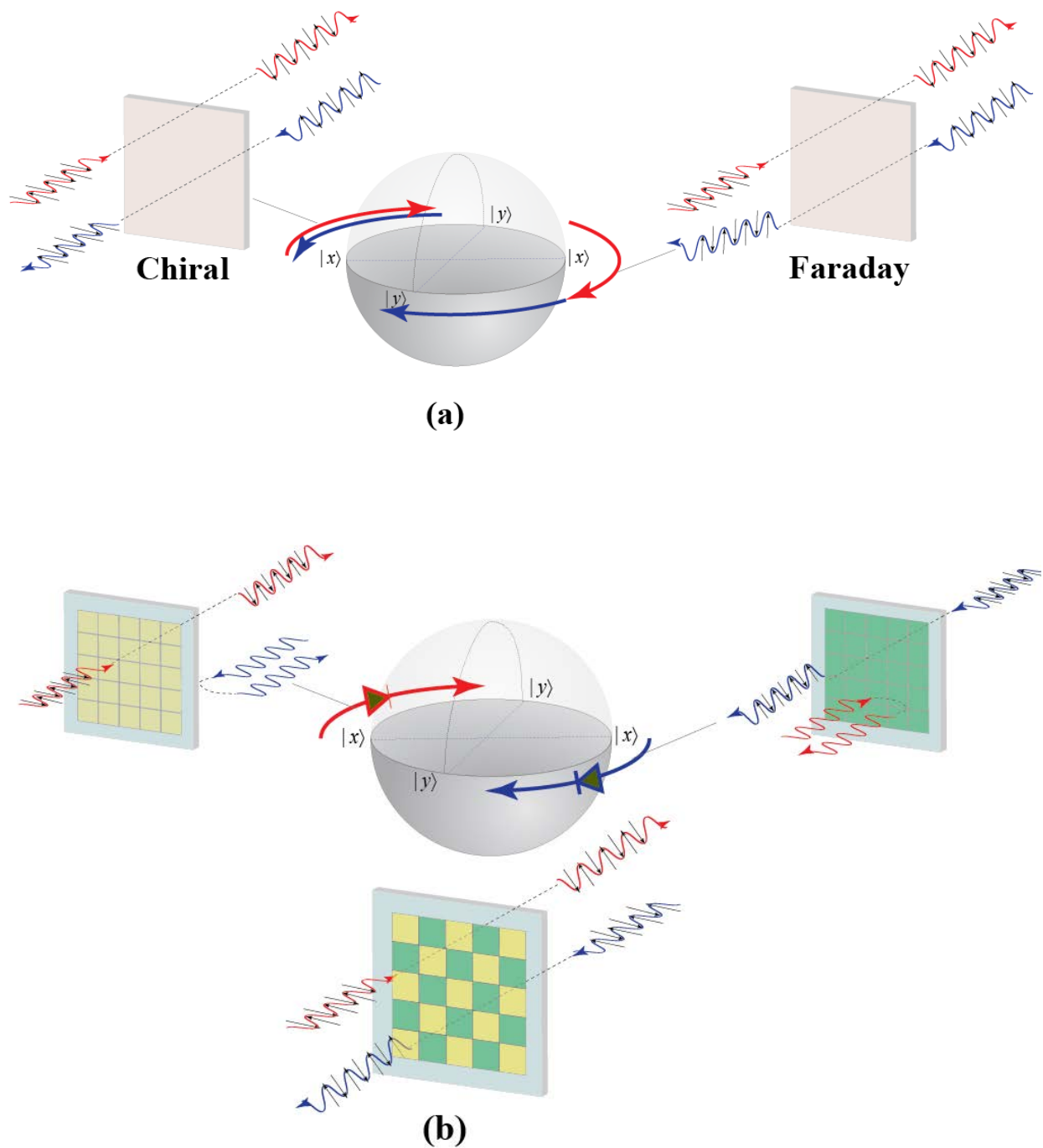

**Fig. 1 Schematics of All-passive nonreciprocal metasurface**: (a) The difference between the 'reciprocal' polarization rotation due to chirality and the 'non-reciprocal' Faraday rotation phenomenon is shown, both mapped onto the Poincaré sphere. (b) Depending on the illumination direction of the incident wave, one of the two constituent

designs (shown as green and yellow), acting as a "wave diode", allows the wave to go through, interacting with the chiral element (not shown) in the unit. Owing to this interaction, the plane of polarization of the wave rotates clockwise by nearly 90 degrees as it goes through this wave diode. The chiral elements in the diodes for waves going into the (+z) direction (light blue design) are mirror image of the chiral elements in the diodes for waves going into the (-z) direction (dark blue design). The all-passive metastructure formed as the checkerboard pattern of such alternating designs may function as a nonreciprocal metasurface mimicking Faraday rotation, while no biasing electric or magnetic field is used.

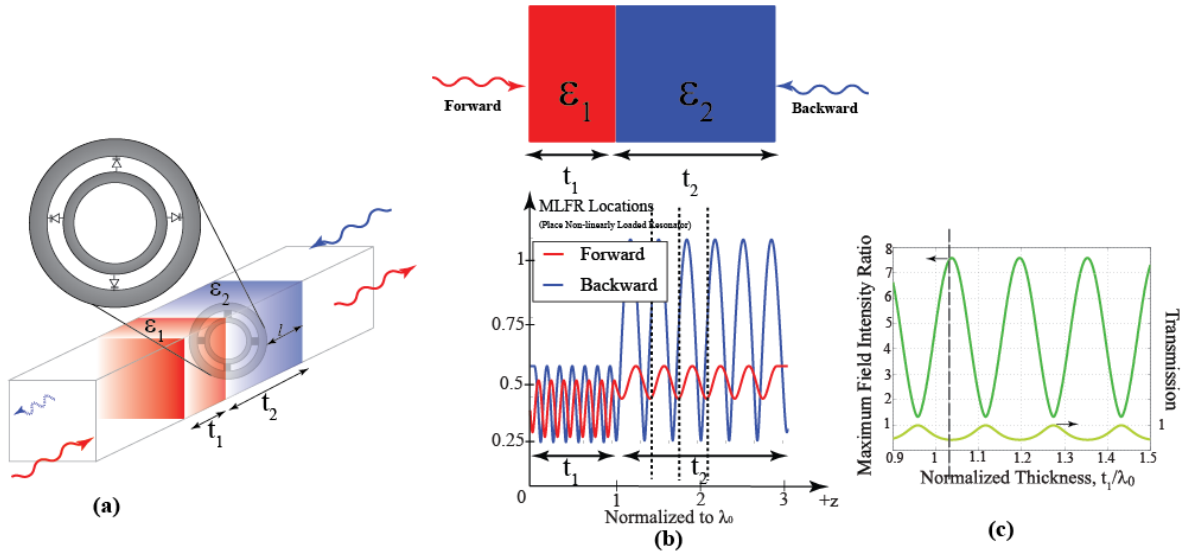

**Fig. 2 Concept of the proposed electromagnetic wave diode**: (a) Bilayered asymmetric slab consisting of two layers with different dielectric permittivities $\varepsilon_1$ and $\varepsilon_2$ and thicknesses $t_1$ and $t_2$. The inset shows the varactor-loaded thin resonator placed at a distance $l$ from one of the slabs ends. The resonator is made of two concentric rings loaded by four varactors distributed symmetrically along its perimeter, (b) The electric field distribution within a bilayered asymmetric slab, The field amplitude distribution within the structure is dependent on the side from which it is excited. The red and blue curves show the field amplitude distribution for the forward (+z) and backward (-z) propagating waves, respectively. The dashed lines show the locations of maximum local field ratio "MLFR", where the ratio between the local field amplitude for the forward and backward illumination is maximum. These are the planes where nonlinearly loaded resonator should be inserted for the most efficient nonlinear response. (c) Maximum ratio of the field intensity (green curve) and transmission coefficient (yellow curve) versus normalized thickness $t_1$. The dashed line shows the operating point used in our design, where both maximum field intensity ratio and transmission coefficient are acceptably high.

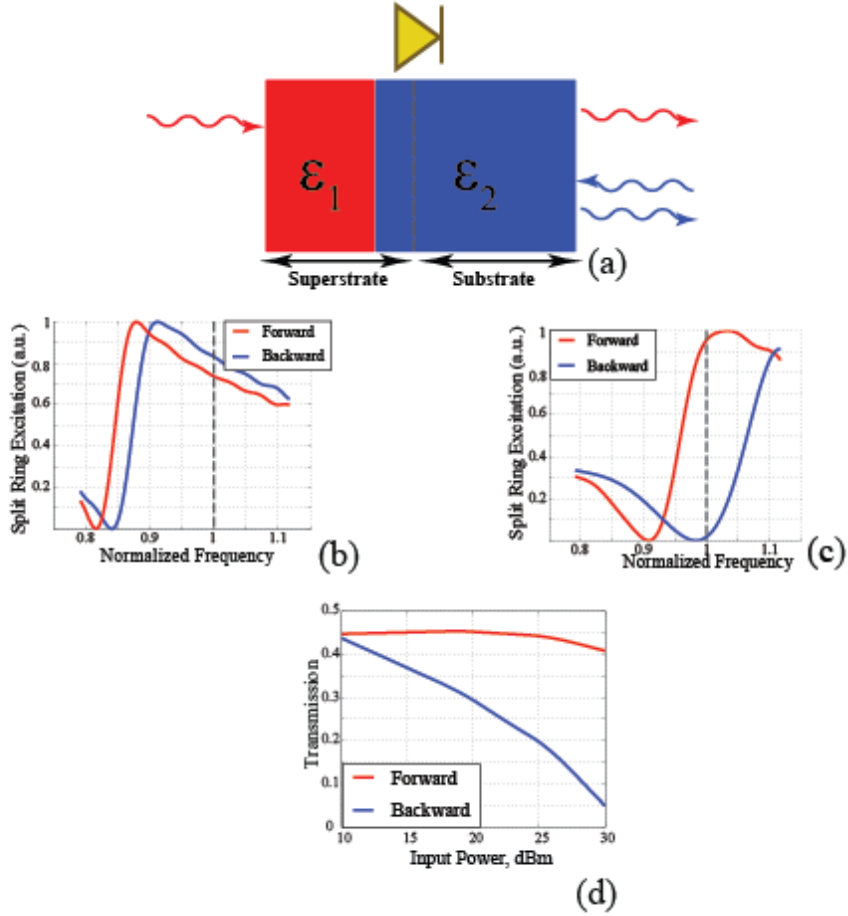

**Fig. 3 Response of the nonlinearly loaded resonator ring and transmission coefficient of electromagnetic wave diode:** (a) Schematics of the wave diode, (b) The ring excitation in arbitrary units for 10 dBm incident power versus the normalized frequency, for both forward (+z) and backward (-z) direction of propagation we get almost the same excitation which yields the same transmission characteristics for both illuminations. (c). The ring excitation in arbitrary units for 30 dBm incident power versus the normalized frequency, the significant difference in the resonance location between forward and backward illumination yields two different transmission coefficients at the operating frequency. (d) Transmission coefficient vs input power level. The red and blue curves show the cases of the forward and backward illuminations, respectively.

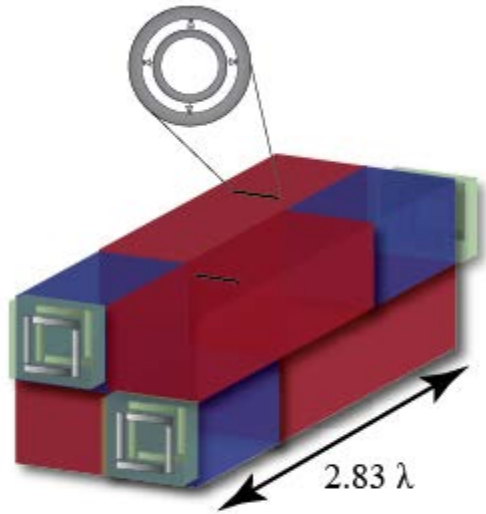

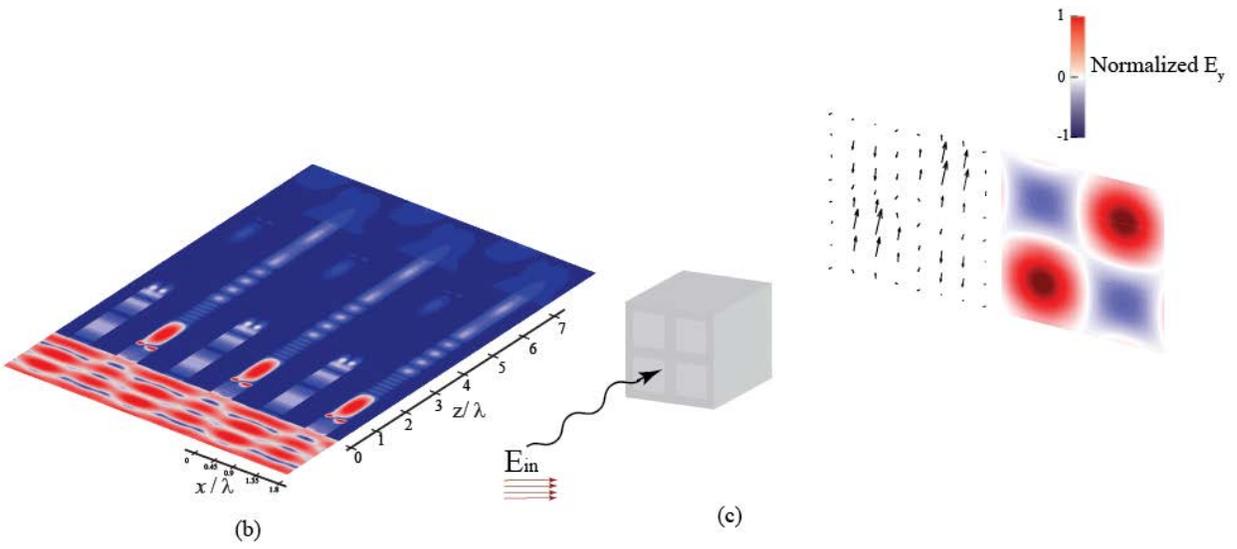

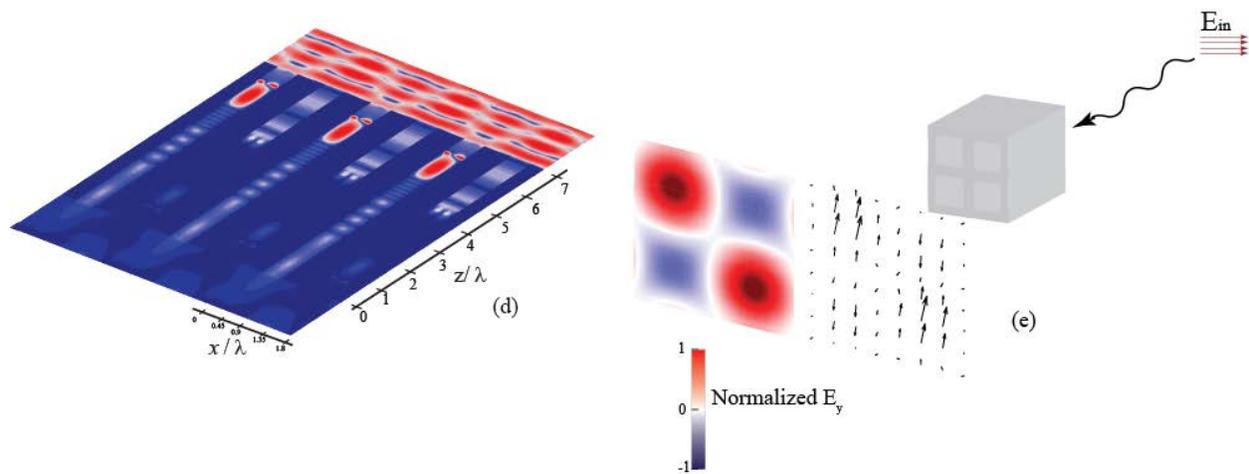

**Fig. 4 Schematic of all-passive quasi-2D nonreciprocal metasurface**: (a) Geometry of the device with four units of the checkerboard including two pairs of mirror-imaged chiral elements shown in Fig. 1. (b) Cross section of the electric field intensity distribution within the metasurface with the incident wave propagating in the +z direction, (c) Simulation results for the output field (shown as compared with the incident field) when the incident wave propagates in the +z direction, (d) and (e) Similar to (b) and (c) but when the incident wave propagates in the -z direction.